\documentclass[12pt]{article}
\usepackage{scicite}
\usepackage{epsfig}

\usepackage{times}

\newenvironment{sciabstract}{%
\begin{quote} \bf}
{\end{quote}}

\newcounter{lastnote}

\title{ Meltdown in quantum computers 
needs not occur: Nuclear experiments 
show a way out}

\author{
J. Flores,$^1$ S. Yu. Kun,$^{1,2,3}$, T.H. Seligman$^1$\\
\\
\normalsize{$^1$ Centro de Ciencias F\'{i}sicas, Universidad Nacional
  Aut\'{o}noma de M\'{e}xico, Cuernavaca, Morelos, Mexico}\\
\\
\normalsize{$^2$ Centre for Nonlinear Physics, RSPhysSE, ANU,
Canberra ACT 0200, Australia}\\
\\
\normalsize{$^3$ Department of Theoretical Physics, RSPhysSE, ANU, 
Canberra ACT 0200, Australia}
}

\date{\today}

\begin{document}
\baselineskip24pt
\maketitle

\begin{sciabstract}
We show that phase memory can be much longer than energy relaxation in 
systems with exponentially large dimensions of Hilbert space; this finding is 
documented by fifty years of nuclear experiments, though the information 
is somewhat hidden. For quantum computers Hilbert spaces of 
dimension $2^{100}$ or larger will be typical and therefore this effect may 
contribute significantly to reduce the problems of scaling of 
quantum computers to a useful number of qubits.
\end{sciabstract}

To solve problems intractable up to now, quantum computers (QC) should 
operate with $n\approx 1000$ interacting qubits. Georgeot and Shepelyansky 
(GS) considered a two-body random Hamiltonian as a generic model for 
QC hardware, 
and performed numerical simulations for $n\leq 15$ [1]. They claim that 
information 
loss, referred to as meltdown of the QC, occurs on a time scale given by 
qubit mixing of eigenstates. Since the dimension of Hilbert space grows 
exponentially and the spectral span only linearly, this poses stringent 
conditions on the interactions among qubits. However, to test these 
restrictions for realistic $n$, we need, alas, a working QC. Instead, we 
resort to old and new nuclear data. We find that using proton-proton 
scattering on heavy nuclei as a quantum protocol, the eigenstate mixing 
time is orders of magnitude shorter than that required for information loss. 
Thus, in exponentially large Hilbert spaces, phase memory, not usually 
considered, is greatly enhanced. Heavy nuclei, therefore, provide a seed 
for a scaling of QC. 

The feasibility of quantum computing on a large scale has been studied from 
different viewpoints. The most common approach is a time dependent one, 
related directly to the increase of errors as a function of the number of 
gates and qubits [2,3]. Fidelity or more specific process-related benchmarks 
are used to get a reliable picture. This approach is self-defeating if one 
wants to scale it to a QC of useful size, and 
simultaneously go beyond perturbation theory [4]. A functioning QC would be 
needed to make the correct calculation with which the perturbed one be 
compared. 

GS point out that, for chaotic dynamics, the 
identity of functions on individual qubits may be lost at a rate faster 
than the quantum protocols [4]. This so called meltdown of the QC would 
put very serious limitations on its implementation. This analysis is based 
on standard theory of relaxation in quantum many-body systems.

The basic assumptions involved are: A qubit is normally a two-level system, 
with an average energy difference $\Delta_0$. For $n$ qubits the level 
density grows 
exponentially with $n$. This, according to GS, imposes stringent restrictions 
on the strength and/or form of the interaction among qubits, since otherwise 
many non-interacting $n$-qubit states $|\Psi_i >$ will be mixed and the 
QC melts 
down. These limitations are particularly damaging since chaotic dynamics 
can stabilise quantum computation external error [5,6].

To investigate parameter values for which the QC can indeed operate, GS 
analysed the statistical properties of the two-body random Hamiltonian   
$H=\sum_iL_i\sigma_i^z+\sum_{i<j}J_{ij}\sigma_i^z\sigma_j^z$, where
where $\sigma_i$ are the Pauli matrices for qubit i. The random numbers 
$L_i$ and $J_{ij}$ 
are distributed respectively in the intervals $[\Delta_0 - \delta /2, \Delta_0 
+ \delta/2]$ and 
$[-J,J]$. Their analysis assumes nearest neighbour coupling.

In  the  non-interacting  qubit  basis  the  eigenfunctions $|\phi >$  are 
obtained. For $n = 12$, $W_i = | <\Psi_i |\phi > |^2$ is plotted as a 
function of the 
non-interacting multi-qubit energy $E_i$ for two values of $J/\Delta_0$. 
For $J/\Delta_0 = 0.02$,
$W_i$ is very narrowly distributed, whereas for $J/\Delta_0 = 0.48$ the 
computer 
eigenstates become a broad and somewhat random mixture of the quantum 
register states $|\Psi_i >$. In the drastic language of GS the meltdown has 
occurred before the quantum protocol could be realised. This implies a time 
scale, which will be introduced below using the standard language of 
statistical nuclear physics.

Wigner, some fifty years ago [7], introduced the spreading width 
$\Gamma^\downarrow$, in the 
context of many-body problems consisting of $n$ interacting  particles,  
with  large  but finite $n$. $\Gamma^\downarrow$ indicates the spread of 
$W_i$ , and  
$\hbar/\Gamma^\downarrow$ is the energy relaxation time for which, 
according to standard theory, 
all memory of the initial state is lost. We shall return to this 
interpretation later. Interchanging the roles of the eigenbasis and the 
single-particle basis, the local density of states (LDOS) is obtained. Its 
width is typically again $\Gamma^\downarrow$.

Unfortunately, for large $n$ it is impossible to perform the calculations of 
GS since the dimension $N_H = 2^n$ of the Hilbert space grows exponentially. 
We 
therefore propose a different approach, using experimental data involving 
heavy nuclei. The nucleus is an ideal laboratory to study many-body systems, 
since nuclear interactions are so strong that external perturbations can be 
neglected.

Consider some scattering process, such as inelastic proton-nucleus scattering 
to be the quantum protocol. The single-particle basis is the quantum register,
the entrance channel represents the loading process, and the output is the 
readout. The question is: How long is the memory and is it given by the 
spreading width or, equivalently, by the width of LDOS?  Since experiments 
of this type have been available for fifty years now [8] and are still 
performed [9], this question can be answered, the nucleus playing 
the role of the QC. 
We address here the phase memory of the process, which is not usually 
considered in the field of compound nuclear reactions, because energy 
relaxation was at the centre of attention.

We revisit the 1954 paper of Gugelot [8] describing the inelastic scattering 
of 18 MeV protons off several targets, including light nuclei such as 
aluminium, medium heavy ones, for example, iron, nickel, copper, silver 
and tin, as well as heavy nuclei such as platinum and gold. The energy 
spectra of the outgoing protons are detected at different angles. The raw 
data are scaled with the proton 
energy $E$ times the penetration factor of the Coulomb barrier to produce 
$I(E)$. At proton energies well below this barrier, where compound reactions 
dominate, the scaled spectra should represent LDOS of the residual nucleus 
and, therefore, be angle independent. This happens for light and medium 
nuclei, as exemplified in Fig. 1 for iron. Surprisingly this is not so for 
heavy nuclei, as shown in Fig. 2 for platinum. The curves are different, 
but the exponential slope at low energies is the same for both angles, 
indicating that energy relaxation has occurred at $\approx 0.7$ MeV per 
proton. 
Gugelot stresses that there are no spurious experimental effects in the 
platinum data, and that gold spectra look similar. In Fig. 3 more recent 
low-energy proton angular distributions obtained from scattering data of 
neutrons [9] and protons [10,11] on a bismuth target, confirm the forward 
peaking. Memory of the direction of the incident beam is clearly retained.

The essential question is: How much time did the protocol, i.e. the reaction  
process,  take  as  compared  to  the energy  relaxation time 
$\hbar / \Gamma^\downarrow$. Using standard nuclear physics estimates [12], 
$\Gamma^\downarrow$  for platinum is 
of the order of 1 MeV. Assuming that we are in a compound state, we can 
estimate the total decay width $\Gamma_{cn}^\uparrow\approx 0.02$ keV 
(see Fig. 7 in Ref. [13]). This 
leads to a process time five orders of magnitude longer than 
$\hbar/ \Gamma^\downarrow$. The 
theoretical estimates given for both widths should not be off by more than 
a factor of three leaving at worst still four orders of magnitude between 
the two time scales. We therefore clearly see that there is strong old and 
new experimental evidence, that $\hbar /\Gamma^\downarrow$ is not the time 
scale for information loss.

Estimates of the effective dimension of Hilbert space can be obtained from 
the spreading width [12] and the density of states [14]. These dimensions and 
the number of qubits needed to roughly equate them, are $10^{20}$ 
($\approx 67$ qubits) 
for $p$ + Pt and $10^9$ ($\approx 30$ qubits) for $p$ + Fe. In view of such 
dimensions, 
digital computations to confirm this effect cannot be performed.

A theoretical explanation of this phase memory persistence in nuclear physics 
is not readily available. There are indications that random two-body 
interactions in exponentially large Hilbert spaces need not lead to chaotic 
states even if all pairs interact [15]. We then could have large spreading 
widths, i.e. strong interaction, but fairly small participation ratios of the 
expansion of one basis in terms of 
the other, as expected for systems with Poissonian statistics in the 
strong-coupling case [16]. This would imply that states are not evenly 
populated, and the average proton energy of 0.7 MeV is then not easily 
explained. We therefore prefer to assume that the time scale for phase 
relaxation is much longer than that for energy relaxation. One of us has 
proposed such ideas some time ago [17,18] showing that very weak correlations 
between different angular momenta may be considerably enhanced in 
exponentially large Hilbert spaces, even if thermalisation occurs for each 
angular momentum. This theory predicts that odd terms in a Legendre 
expansion of the angular distribution will not vanish, but be determined 
by the ratio of decay time and phase relaxation time [19]. The corresponding 
fits are shown in Fig. 3 and the two time scales agree. This is consistent 
with our statement that phase relaxation is orders of magnitude slower than 
energy relaxation.

The good news is then that there are fifty years of strong experimental 
evidence that the energy relaxation time is not the relevant time scale 
that limits memory conservation in a system of many qubits. We have 
identified an effect, observable only in exponentially large Hilbert spaces, 
that introduces a much longer time scale for phase memory in a many-body 
system. In principle, this effect allows scaling to large number of qubits, 
although it will certainly not replace stabilisation techniques developed for 
small-$n$ QC [2,3]. The bad news is that we need more theoretical insight for 
appropriate engineering of a QC to take advantage of this effect. 

We are grateful to Dr. E. Raeymackers for making available to us tables of 
the data reported in Ref. [9]. We thank T. Guhr, T. Prosen and 
F. Leyvraz for useful discussions. This work was supported by Direcci\'on 
General de Asuntos de Personal Acad\'emico, Universidad Nacional Aut\'onoma 
de M\'exico, project IN101603 and by Consejo Nacional de Ciencia y 
Tecnolog\'ia, Mexico, project 41000 F.

\vskip2cm
\noindent
{\bf References}
\bibliographystyle{Science}

\noindent
[1] B. Georgeot, D.L. Shepelyansky, {\sl Phys. Rev. E} {\bf 62} 3504 (2000).

\noindent
[2] M.A. Nielsen, I.L. Chuang, {\sl Quantum Computation and Quantum 
Information} (Cambridge, 2000).

\noindent
[3] G.P. Berman, G. Doolen, R. Maineri, V.I. 
Tsifrinovitch, {\sl Introduction to Quantum Computers} (World Scientific, 
1998).

\noindent
[4] G.P. Berman, G.D. Doolen, D.I. Kamenev, V.I. Tsifrinovich, 
{\sl Phys. Rev. A} {\bf 65} 012321 (2002).

\noindent
[5] T. Prosen, M. Znidaric, {\sl J. Phys. A: Math. Gen.} {\bf 34} L681 (2001).

\noindent
[6] G.L. Celardo, C. Pineda, M. Znidaric, {\sl Stability of quantum Fourier 
transformation on Ising quantum computer}, LANL e-print 
arXiv: quant-ph/0310163.

\noindent
[7] E.P. Wigner, {\sl Ann. Math.} {\bf 62} 548 (1955); {\bf 65} 203 (1957).

\noindent
[8] P.C. Gugelot, P.C. {\sl Phys. Rev.} {\bf 93} 425 (1954).

\noindent
[9] E. Raeymackers {\sl et al}, {\sl Nucl. Phys. A} {\bf 726} 210 
(2003). 

\noindent
[10] F.E. Bertrand, R.W. Peelle, {\sl  Phys. Rev. C} {\bf 8} 
1045 (1973).

\noindent
[11] F.E. Bertrand, R.W. Peelle, ``Cross sections of hydrogen 
and helium particles produced by 62-and 39-MeV protons on $^{209}$Bi''. 
(Oak Ridge National Laboratory, Report No. 4638, 1971).

\noindent
[12] D. Agassi, H.A. Weidenmüller, G. Mantzouranis, {\sl Phys. Rep.} 
{\bf 22} 145 (1975).

\noindent
[13] T. Ericson, T. Mayer-Kuckuk, {\sl Ann. Rev. Nucl. Sci.} {\bf 16} 183 
(1966).

\noindent
[14] A. Bohr, B.R. Mottelson, {\sl  Nuclear Structure}, (Benjamin, New York, 
1969), Vol. 1, p. 284.

\noindent
[15] L. Benet, H.A. Weidenmüller, {\sl J. Phys. A: Math. Gen.} {\bf 36} 
3569 (2003).

\noindent
[16] L. Benet, T.H. Seligman, H.A. Weidenmüller, {\sl Phys. Rev. Lett.} 
{\bf 71} 529 (1993).

\noindent
[17] S.Yu. Kun, {\sl  Z. Phys. A} {\bf 348} 273 (1994).

\noindent
[18] S.Yu. Kun, {\sl  Z. Phys. A} {\bf 357} 255 (1997).

\noindent
[19] S.Yu. Kun, A.V. Vagov, A. Marcinkowski, {\sl Z. Phys. A} {\bf 358} 69 
(1997).

\clearpage

\noindent
{\bf Figure captions}

\noindent {\bf Fig. 1.} Scaled proton spectra $I(E)$ (in arbitrary units) at
forward and backward angles for 18 MeV proton inelastic scattering on iron
(reproduced from Fig. 3 of Ref. [8]). They represent relative LDOS of the
residual nucleus for high excitation energy, i.e. low proton energy.

\noindent {\bf Fig. 2.} Similar spectra as in Fig. 1 for a platinum target 
(reproduced from Fig. 9 of Ref. [8]). Note that the vertical scale does no
longer represent LDOS in any range because of the significant difference
between backward and forward angles.

\noindent {\bf Fig. 3.} Angular distribution (dots) of inelastic sub-Coulomb 
9 MeV protons measured with a 62 MeV beam on a bismuth target [10,11]. Circles
represent a similar distribution of $9\pm 1$ MeV protons resulting from
$62.7\pm 2$ MeV neutron induced reactions on bismuth [9]. The full lines are
fits with Legendre polynomials up to second order.

\clearpage

\noindent
\begin{figure}
\includegraphics{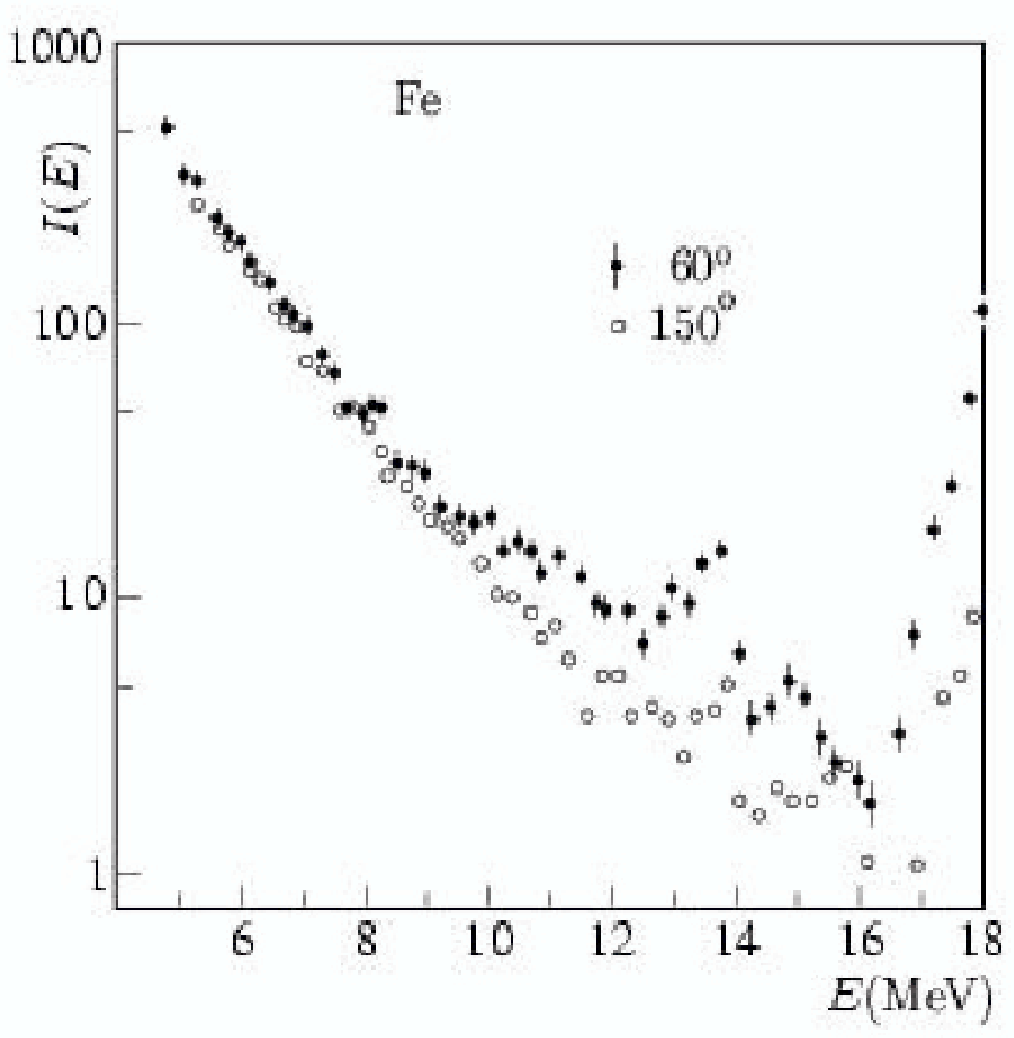}\\
\caption{ }
\end{figure}

\begin{figure}
\includegraphics{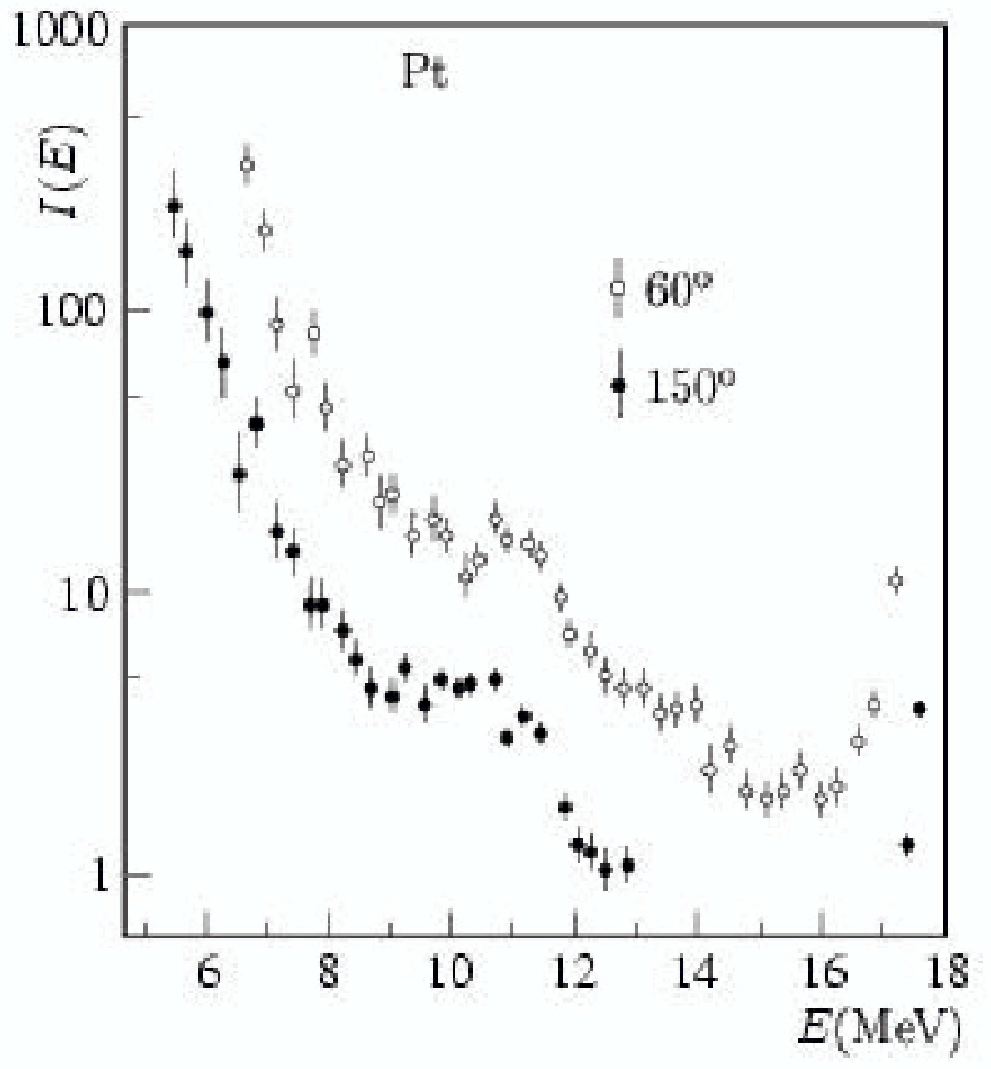}\\
\caption{}
\end{figure}

\begin{figure}
\includegraphics[width=11cm,angle=90]{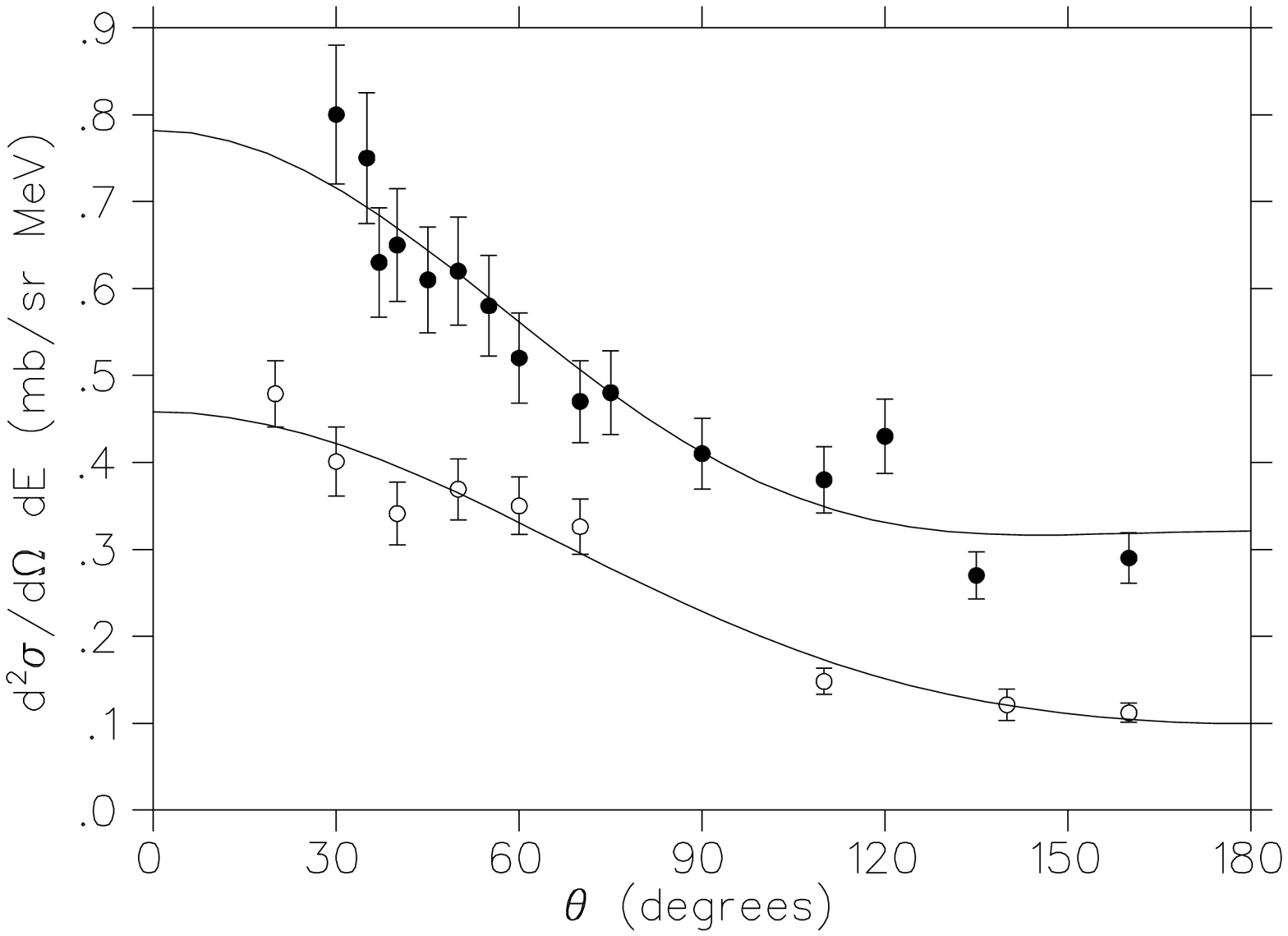}\\
\caption{ }
\end{figure}

\end{document}